\documentclass[12pt,preprint]{emulateapj}
\usepackage{graphicx}
\usepackage{epstopdf}
\usepackage{color}
\def\apj{{ApJ}}
\def\apjl{{ApJL}}
\def\mnras{{ MNRAS}}

\def\nat{{Natur}}
\def\GCN{{ GCN Circ}}

\def\be{\begin{equation}}
\def\ee{\end{equation}}
\def\bea{\begin{eqnarray}}
\def\eea{\end{eqnarray}}

\begin{document}
\title{The interpretation of the multi-wavelength afterglow emission of short GRB 140903A}
\author{Shuai Zhang$^{1,2}$, Zhi-Ping Jin$^{1}$, Yuan-Zhu Wang$^{1,2}$, and Da-Ming Wei$^{1}$}
\affil{$^1$ Key Laboratory of Dark Matter and Space Astronomy, Purple Mountain Observatory, Chinese Academy of Sciences, Nanjing, 210008, China.}
\affil{$^2$ University of Chinese Academy of Sciences, Yuquan Road 19, Beijing, 100049, China.}
\email{jin@pmo.ac.cn (ZPJ), dmwei@pmo.ac.cn (DMW)}

\begin{abstract}
GRB 140903A, a short duration $\gamma-$ray burst (SGRB) detected by {\it Swift}, is characterized by its long-lasting radio emission among SGRBs. In addition to the $\sim 10^{6}$ s radio afterglow emission, the afterglow of GRB 140903A displays a plateau from $10^3$ s to $7 \times 10^3 $ s in the X-rays. In this work, we attribute the X-ray plateau to the energy injection into the decelerating blast wave and then model the later radio/optical/X-ray afterglow emission within the standard fireball afterglow model. The afterglow emission has been well reproduced with reasonable physical parameters, including a jet half-opening angle $\sim 0.05$. We also compare the physical properties of SGRBs with and without radio afterglow detections and find that nearby SGRBs are brighter and easier to be detected in radio. Such a phenomena has been interpreted within the afterglow model.
\end{abstract}

\section{Introduction}
Gamma-ray bursts (GRBs) are the most energetic flashes of soft $\gamma-$ray from the deep universe. Based on their duration distribution, GRBs can be generally divided into two groups, including the so-called short GRBs (SGRBs; the duration is shorter than 2 second) and long GRBs. \citep{Kouveliotou1993}. Usually the long GRBs are associated with bright supernovae and thus should be from massive star collapse \citep{Woosley2006}. The SGRBs, however, are expected to be from neutron star mergers (including both the double neutron star mergers and neutron star-black hole mergers) \citep{Eichler1989}. The compact object merger origin of SGRBs has been strongly supported by their afterglow and host galaxy observations \citep{Berger2014ARA&A, Fong2015} and the identification of Li-Paczynski macronovae \citep{Li1998, Kasen2013} in SGRB 130603B \citep{Tanvir2013, Berger2013}, long-short GRB 060614 \citep{Yang2015,Jin2015} and SGRB 050709 \citep{Jin2016}. SGRBs are hence widely believed to be promising gravitational wave (GW) sources and important sites of generating heavy elements \citep{Eichler1989, Jin2016} and the SGRB/GW association is expected to be reliably established in the advanced LIGO/Virgo era (see \citet{Li2016} and the references therein). The observations of SGRBs have attracted wider and wider attention and the modeling of the afterglow can provide additional information. For instance the derivation of the half-opening angle of the SGRB ejecta is important for estimating the detection prospect of GW emission from neutron star mergers by advanced LIGO/Virgo network \citep{Berger2014ARA&A} and the afterglow modeling may reveal the nature of the central engine of short GRBs.

In this work we focus on GRB 140903A, a short burst characterized by its long-lasting radio emission. Before the detection of this event, there had been just three short GRBs with radio afterglow  and all had been detected by The Jansky Very Large Array (VLA), including GRB\,050724A \citep{Berger2015,pan06}, GRB\,051221A \citep{sbk+06} and GRB\,130603B \citep{Fong2014}. Among the limited sample, GRB 140903A has the longest radio afterglow detection ($\sim 10^6 $ s). In addition to the long-lasting radio afterglow emission, the afterglow of GRB 140903A displays a plateau from $10^3$ s to $7 \times 10^3$ s in the X-rays. The main purpose of this work is to interpret these features. We also carry out the statistical study to examine why the radio afterglow detection of SGRBs are so rare. 

Recently \citet{Troja2016} reported their observing result of GRB 140903A. They monitored the X-ray afterglow up to 15 days with {\it Chandra}. Together with the optical and radio afterglow data, they found an achromatic jet break at about 1 day after the burst. They discussed the nature of the GRB progenitor system and concluded that this event likely originated from a compact binary merger. An off-axis jet model was used to interpret the multi-band afterglow. In this work we carry out an independent analysis in the on-axis jet scenario.

\section{The afterglow emission of SGRB 140903A and the interpretation}
\subsection{The afterglow data}

At 15:00:30 UT on 2014 Oct 03, GRB 140903A triggered the Burst Alert Telescope (BAT) onboard the {\it Swift} satellite \citep{Cummings2014}.Its duration $T_{90}$ (15-350 keV) is $0.30 \pm 0.03$ second. The time-averaged spectrum between ${\rm T}_{0}-0.01$ to ${\rm T}_{0}+0.35$ second is best fitted by a simple power-law model, with a photon index $1.99 \pm 0.12$ \citep{Palmer2014}.

{\it Swift} XRT began to observe the position of GRB 140903A at about 59 seconds after the BAT trigger \citep{Kennea2014} and detected the afterglow \citep{DePasquale2014}.
The X-ray afterglow decay firstly ($\alpha\sim0.2$) before 1000 seconds and then followed by a plateau with a flux $ \sim 10^{-11} {\rm erg}~{\rm cm}^{-2}~{\rm s}^{-1}  $ until about $10^4$ s and then turned to a more rapid decay \citep{DePasquale2014,Troja2016}. {\it Chandra} took two epochs of observations of the afterglow at about 3 and 21 days after the burst and both detected the source \citep{Sakamoto2014,Troja2016}.
Compared them with the {\it Swift} XRT data, there was a jet break at about 1 days \citep{Troja2016}.

The afterglow of GRB 140903 had been observed by many ground-based telescopes. Among them, the 4.3m Discovery Channel Telescope \citep{Capone2014}, the robotic Palomar 60 inch telescope \citep{Cenko2014}, the Gemini-North telescope  \citep{Cucchiara2014,Levan2014}, the Pan STARRS survey telescope \citep{Fruchter2014}, the Faulkes Telescope North \citep{Dichiara2014} and the Nordic Optical Telescope \citep{Xu2014} detected the optical counterpart of the X-ray afterglow. Gemini GMOS observed the source in spectroscopic mode at about 14.6 hours after the burst, and the redshift $z=0.351$ was determined by identifying the H$\beta$ and OIII emission lines and the NaI absorption line \citep{Cucchiara2014}. This was confirmed by later spectroscopy of the host galaxy by GTC, which detected a H$\alpha$ line at the same redshift \citep{Troja2016}.
The radio afterglow of 6.1 GHz and 9.8 GHz was also detected by the Very Large Array (VLA) at about 10 hours after the trigger of BAT \citep{Fong2014GCN}, it was still detectable until $10^6$ s. The radio flux rise firstly and then decay, the peak flux density is $\sim 200 \mu \rm Jy $ at about $ 2 \times 10^5 $ s \citep{Troja2016}.
The Giant Metrewave Radio Telescope (GMRT) also detected the radio emission \citep{Nayana2014}.
In this work we give a interpretation of these observational features within the framework of external forward shock model with energy injection.

\subsection{X-ray afterglow flat segment due to the energy injection}\label{sec:Theor}

Flat segment of X-ray afterglow has been detected in a good fraction of long GRB afterglows \citep{Nousek2006, Zhang2006} and some SGRB afterglows \citep{Burrows2006}. Such a kind of phenomena has been widely interpreted as the energy injection from the prolonged activity of the central engine \citep{Zhang2006, Fan2006}. However, sometimes the optical afterglow emission can not be interpreted self-consistently within such a scenario \citep{Fan2006b,Liang2007,Panaitescu2006b}.

The energy injection can be written generally as \citep{Cohen1999, Zhang2001, Zhang2006}:
\begin{equation}
\label{eninj}
\frac{dE}{dt}=A(1+z)^{-1}\Big(\frac{t}{t_{\rm o}}\Big)^{-q},~~t_{\rm i}<t<t_{\rm f}
\end{equation}
where $t_{\rm i}$ and $t_{\rm f}$ represent the start and end time of energy injection. At the time of $t_{\rm c}$ (when the amount of injected energy equals to the initial kinetic energy in the outflow) the dynamical evolution of the GRB ejecta is changed significantly, i.e. $ \int_{t_{\rm i}}^{t_{\rm c}}(dE/dt)dt \sim E_{\rm k} $, where $E_{k}$ is the kinetic energy of the outflow. Then we get $ A t_{\rm o}^q(t_{\rm c}^{1-q}-t_{\rm i}^{1-q}) \sim (1+z)(1-q)E_{\rm k} $.

From observations we know that $ \nu_{\rm c}(t_{\rm d}=1) \gtrsim \rm 10 ~keV $ for SGRB 140903A. In this case, the X-ray flat plateau decline as $ \alpha=\frac{(2p-4)+(p+2)q}{4} $  \citep{Zhang2006}. For typical value of $p \sim 2.2-2.5$, the X-ray flat segment of SGRB 140903A is in agreement with the case of $q \sim 0$, i.e., the central engine is a pulsar-like compact object. Similar conclusions have been drawn for example for SGRB 151221A \citep{Fan2006}. In the case of long GRBs, please see \citet{Dai1998} and \citet{Zhang2001}. For $q\sim0$, we have $ A \sim \frac{(1+z)E_{\rm k}}{t_{\rm c}-t_{\rm i}}$.

We consider the energy injection \citep{Pacini1967, Gunn1969} as:
\begin{equation}
\label{ejr}
\frac{dE}{dt} \approx \frac{ 6.2\times 10^{47} }{1+z}~\rm erg~\rm s^{-1} B_{\perp,14}^{2}R_{\rm s,6}^{6}\Omega_4^{4}\Big[1+\frac{t}{(1+z)T_{\rm o}}\Big]^{-2},
\end{equation}
where $B_\perp$  and $R_{\rm s}$ are the component of the dipole magnetic field strength perpendicular to the rotation axis and radius of the magnetar respectively, $\Omega$ is the initial angular frequency of rotation. The initial spin-down time-scale in the rest frame is:
\begin{equation}
\label{sdtime}
T_{\rm o} \approx 1.6 \times10^5 \rm s~B_{\perp,14}^{-2}\Omega_4^{-2}I_{45}R_{\rm s,6}^{-6}
\end{equation}
and in which $I \sim 10^{45}$ is the typical moment of inertia of magnetar.
Here and after, we adopted the convention $Q_x=Q/10^x$ in cgs units.

When $t\ll (1+z)T_o$, the energy injection rate is a constant, so with our magnetar model one requires $q\sim 0$ for $t_{\rm c}<(1+z)T_{\rm o}$, we can derive:
\begin{equation}
\label{A}
A \sim \frac{(1+z)E_{\rm k}}{t_{\rm c}-t_{\rm i}} \sim 6.2\times 10^{47}~\rm erg~s^{-1} B_{\perp,14}^{2}R_{\rm s,6}^{6}\Omega_4^{4}
\end{equation}

For $t>(1+z)T_{\rm o}$, the energy injection rate drops rapidly. So we have
\begin{equation}
\label{tf}
t_{\rm f} \sim (1+z)T_{\rm o} \sim 1.6\times10^5 \rm s ~(1+z)B_{\perp,14}^{-2}\Omega_4^{-2}I_{45}R_{\rm s,6}^{-6}
\end{equation}

The total energy injection can be estimated as $ E_{\rm inj}=A T_{\rm o} $, so together with (\ref{sdtime}) and (\ref{A}) we can get $ E_{\rm inj}=t_{\rm f} E_{\rm k} /(t_{\rm c}-t_{\rm i}) \sim 10^{53} ~\rm erg~ I_{45} \Omega_4^2$. Then the ratio of total injected energy to initial kinetic energy is:
\begin{equation}
\label{f}
\frac{E_{inj}}{E_{\rm k}} \sim \frac{ 10^{53} \rm erg~I_{45}\Omega_4^2}{E_{\rm k}}
\end{equation}

From the X-ray afterglow of GRB 140903A, we can infer that the energy injection begins important during about $1000 \sim 2000$ s and ends at about $7000$ s, so we estimate $ t_{\rm c}-t_{\rm i} \sim 1000 $ s and $ t_{\rm f}\sim 7000 $ s. So $E_{\rm inj}/E_{\rm k} \sim t_{\rm f} /(t_{\rm c}-t_{\rm i})\sim 7 $.  Substituting this into equation (\ref{f}) we get $E_{\rm k} \sim 1.4\times 10^{52} ~\rm erg~ I_{45} \Omega_4^2$. Then by equation (\ref{A}) we obtain $ A \sim   1.9 \times 10^{49} ~\rm erg~ I_{45} \Omega_4^2$, here $ z=0.351 $ is adopted. By assuming typical values of $ I_{45}=1 $ and $ \Omega_4=0.58 $ \citep{Shapiro1983book}, we can estimate $ E_{\rm k} \sim 4.7 \times 10^{51} ~\rm erg $ and $ A \sim 6.4 \times 10^{48} ~\rm erg~\rm s^{-1} $.

We can also constrain the ellipticity $\epsilon \equiv \frac{a-b}{(a+b)/2}$ of the magnetar if we assume gravitational wave radiation is not important. The spin-down time-scale of gravitational wave radiation can be estimated as $ \tau_{\rm gw} \sim 3\times 10^{-3} \epsilon^{-2} I_{45}^{-1} \Omega_4^{-4}$ s \citep{Shapiro1983book}. From $ \tau_{\rm gw} >t_{\rm f} \sim 7000 $ s we can get $ \epsilon < 6.5\times10^{-4}~I_{45}^{-1/2} \Omega_4^{-2}$.

\subsection{Analytical estimate of forward shock physical parameters}\label{sec:PhPa}

The forward shock emission from GRB outflow can be parameterized as \citep[e.g.,][]{Piran1999,Yost2003,Fan2006b}:
\begin{eqnarray}
&& F_{\nu,{\rm max}} = 6.6~{\rm mJy}~\Big({1+z\over 2}\Big) D_{L,28.34}^{-2}\epsilon_{B,-2}^{1/2}E_{k,53}n_0^{1/2}, \label{eq:F_nu,max}\\
&& \nu_{\rm m} = 2.4\times 10^{16}~{\rm Hz}~E_{\rm k,53}^{1/2}\epsilon_{\rm B,-2}^{1/2}\epsilon_{e,-1}^2 C_{\rm p}^2 \Big({1+z \over 2}\Big)^{1/2} t_{\rm d,-3}^{-3/2},\label{eq:nu_m}\\
&& \nu_{\rm c} = 4.4\times 10^{16}~{\rm Hz}~E_{\rm k,53}^{-1/2}\epsilon_{B,-2}^{-3/2}n_0^{-1} \Big({1+z \over 2}\Big)^{-1/2}t_{\rm d,-3}^{-1/2}{1\over (1+Y)^2}, \label{eq:nu_c}
\end{eqnarray}
where $C_{\rm p} \equiv 13(p-2)/[3(p-1)]$, $\epsilon_{\rm e}$ ($\epsilon_{\rm B}$) is the fraction of energy of shocked electrons (magnetic field),  the Compton parameter $Y\sim
(-1+\sqrt{1+4\eta \epsilon_{\rm e}/\epsilon_{\rm B}})/2$, $\eta \sim \min\{1,(\nu_{\rm m}/\bar{\nu}_{\rm c})^{(p-2)/2} \}$ and $\bar{\nu}_{\rm c}=(1+Y)^2 \nu_{\rm c}$.

Equations (\ref{eq:F_nu,max}) to (\ref{eq:nu_c}) can be transferred to \citep{Zhang2015}:

\begin{eqnarray}
\epsilon_{\rm B,-2}^{1/2}E_{\rm k,53}n_0^{1/2}&\approx a,\label{a}\\
E_{\rm k,53}^{1/2}\epsilon_{\rm B,-2}^{1/2}\epsilon_{\rm e,-1}^2 &\approx b,\label{b}\\
E_{\rm k,53}^{-1/2}\epsilon_{\rm B,-2}^{-3/2}n_0^{-1}{(1+Y)^{-2}}&\approx c.\label{c}
\end{eqnarray}

where:
\begin{eqnarray}
a=&&\frac{1}{6.6}{~\rm mJy^{-1}}F_{\nu,{\rm max}} D_{L,28.34}^{2}\Big({1+z\over 2}\Big)^{-1}\\
b=&&\frac{1}{2.4}\times 10^{-16}{~\rm Hz^{-1}}\nu_{\rm m} C_{\rm p}^{-2} \Big({1+z \over 2}\Big)^{-1/2}t_{\rm d,-3}^{3/2}\\
c=&&\frac{1}{4.4}\times 10^{-16}{~\rm Hz^{-1}}\nu_{\rm c}\Big({1+z \over 2}\Big)^{1/2}t_{\rm d,-3}^{1/2}
\end{eqnarray}

\begin{figure}
\begin{center}
\includegraphics[width = 250pt]{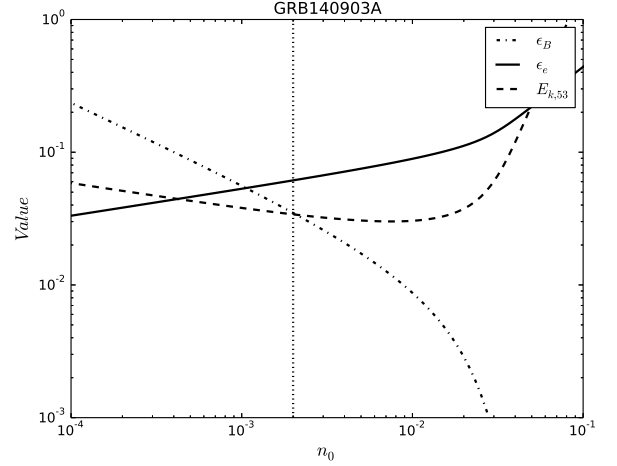}
\end{center}
	\caption{Numerical solutions of equations of (\ref{a}) to (\ref{c}) as functions of $n_0$ under assumption of $ \nu_{\rm c}(t_{\rm d}=1) \sim 10\rm ~keV $. The vertical dotted line shows the location of $n_0=0.002$. The dot-dashed, solid and dashed lines represent the constraints on $\epsilon_{\rm B}$, $\epsilon_{\rm e}$ and $E_{\rm k}$, respectively.}
\label{fig:par}

\end{figure}

For observations we know that $ z=0.351 $, $ p \approx 2.45 $, $ F_{\nu,{\rm max}} \approx 180~ \mu \rm Jy $, $ \nu_{\rm m}(t_{\rm d}=2.5)= 37 ~\rm GHz$ \citep{Troja2016} and $ \nu_{\rm c}(t_{\rm d}=1) \gtrsim \rm 10 ~keV $. Then we can solve equations (\ref{a}) to (\ref{c}) numerically to obtain the  values of $\epsilon_{\rm B}$ , $\epsilon_{\rm e}$ and $E_{\rm k}$ as a functions of $n_0$, which are shown in Fig. \ref{fig:par}. Since we only know the lower-limit of $ \nu_{\rm c}$, we can only obtain the upper-limit of $\epsilon_{\rm B}$ and lower limits of $\epsilon_{\rm e}$ and $E_{\rm k}$ for a fixed value of $n_0$. For example, when $n_0=0.002$, we have $\epsilon_{\rm B} \lesssim 0.03$, $\epsilon_{\rm e} \gtrsim 0.06$ and $E_{\rm k} \gtrsim 3.4\times10^{51}~\rm erg$. Usually, $\max\{\epsilon_{\rm e},~\epsilon_{\rm B}\}\leq 1/3$ (the equipartition assumption) is required and we need $n_0<0.08$. Interestingly, this is in agreement with the expectation that the compact object mergers usually take place in lower density regions.

\subsection{Numerical fit to the data}\label{sec:Num}
\begin{figure}
\begin{center}
\includegraphics[width = 250pt]{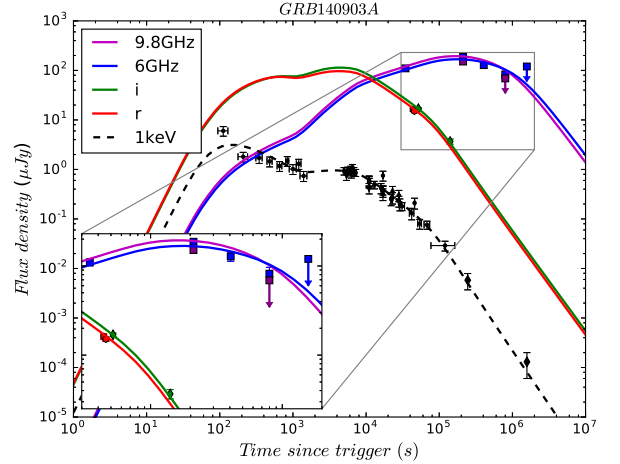}
\end{center}
	\caption{Fit of the multi-band afterglow light-curve. Black, red, green, blue and purple represent X-ray (1keV), r, i, 6GHz and 9.8GHz bands respectively. The inset shows the fit of optical and radio observations.}
\label{fig:fits}
\end{figure}

We fitted the multi-band afterglow numerically, the code was introduced in \citet{Fan2006b}.
If we don't consider the optical extinction, the radio, $i$-band and X-ray data can be fitted well but $r$-band is not acceptable (the reduced total chi-square is $ \chi^2/dof \sim 3.05 $ while $r$-band contribute almost half of it.), therefor extinction must be considered. Troja et al. studied the afterglow spectral energy distribution of GRB 140903A and gave the Galactic extinction $E_{\rm B-V} \approx 0.03$ and an intrinsic extinction $E_{\rm B-V} \approx 0.15$ \citep{Troja2016}. Assuming the host extinction model is the same as the Galactic, we can get the optical extinction $A_r \approx 0.68$ and $A_{\rm i} \approx 0.5$.

Considering this extinction, we can fit the radio, optical and X-ray light curves well. We use the scenario of a normal expanding jet with a later energy injection.
The numerical results are presented in Fig.\ref{fig:fits}, we found this set of fitting parameters $(\epsilon_{\rm e},~\epsilon_{\rm B},~p,~n_{0},~E_{\rm k,51},~\theta_{\rm j},~t_{\rm i},~t_{\rm f},~A)\sim(0.06, ~0.01, ~2.45, ~0.002, ~5, ~0.05, ~1000, ~7000, 6\times10^{48})$ can reasonably fit the data, where $\theta_{\rm j}$ is the half-opening angle of the GRB outflow. Due to the lack of afterglow data in the early time, the only constraint on the initial Lorentz factor is $ \Gamma_{0} \geq 200 $, and we took 200 in the fit. The values of $ E_{\rm k,51} $ ,$ t_{\rm i} $, $ t_{\rm f} $ and $ A $ are consistent with previous analytical results. The reduced chi-square is $ \chi^2/dof \sim 1.3$.

In our fit (see Fig. \ref{fig:fits}), the X-ray afterglow first decay before $ ~1\times10^3 $ s and followed by a continuous energy injection till $ ~7\times10^3 $ s and then turn to a normal decay. At about $ ~1\times10^5 $ s a jet break appeared.

In a previous work, \citet{Zhu2015A&A} have studied another short GRB 130912A which also has a long-lasting optical plateau. However, in that case, the plateau phase can be explained without an energy injection. Here, GRB 140903A has a longer-lasting X-ray plateau and this could be explained by energy injection from $ ~10^3 $ s to $ ~7\times10^3 $ s.

\citet{Troja2016} got a different set of physical parameters $(\epsilon_{\rm e},~\epsilon_{\rm B}, ~n_0,~E_{\rm k,iso},~\theta_{\rm j})\sim(~0.14_{-0.06}^{+0.19}, ~2.1_{-1.4}^{+3.6}\times 10^{-4}, ~0.032_{-0.026}^{+0.14}, ~4.3_{-2.0}^{+1.2}\times 10^{52}, ~0.090 \pm 0.012 \rm ~rad)$ and an observing angle of $~\theta_{\rm obs} \sim 0.055 \rm ~rad $. This is natural since we use the energy injection model which is different from their off-axis jet model. We note that the difference in density may explain most of the differences in the physical parameters. However, one should note that in our energy injection model, the total kinetic energy at the end of energy injection is $\sim 4.1\times10^{52}$ erg, which is consistent with that of \citet{Troja2016}.

\section{Statistical study}

\begin{figure}
\begin{center}
\includegraphics[width = 250pt]{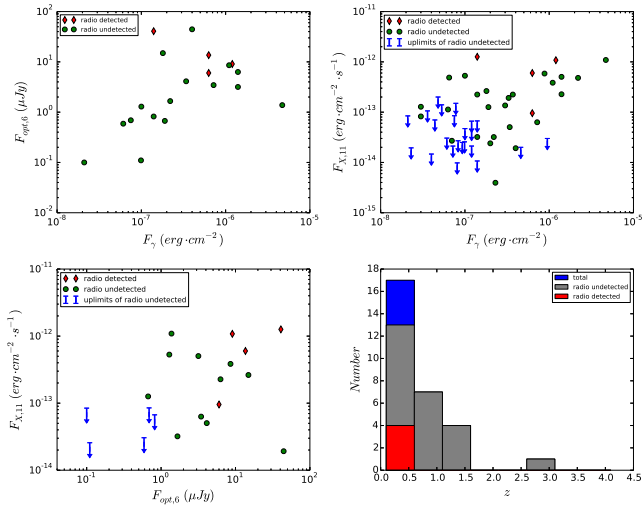}
\end{center}
\caption{ \emph{Upper left:}  Relation between $F_{\rm opt,6}$ and $F_{\gamma}$.  \emph{Upper right:} Relation between $F_{\rm X,11}$ and $F_{\gamma}$. \emph{Lower left:} Relation between $F_{\rm X,11}$ and $F_{\rm opt,6}$. \emph{Lower right:} Distribution of redshift. Optical and X bands flux densities are observed at 6 and 11 hours after bursts. Green dots and blue upper-limits represent GRBs without radio afterglow detections and red dots represent GRBs with radio afterglow detections.}
\label{fig:relations}
\end{figure}

\begin{figure*}
\begin{center}
\includegraphics[width = 450pt]{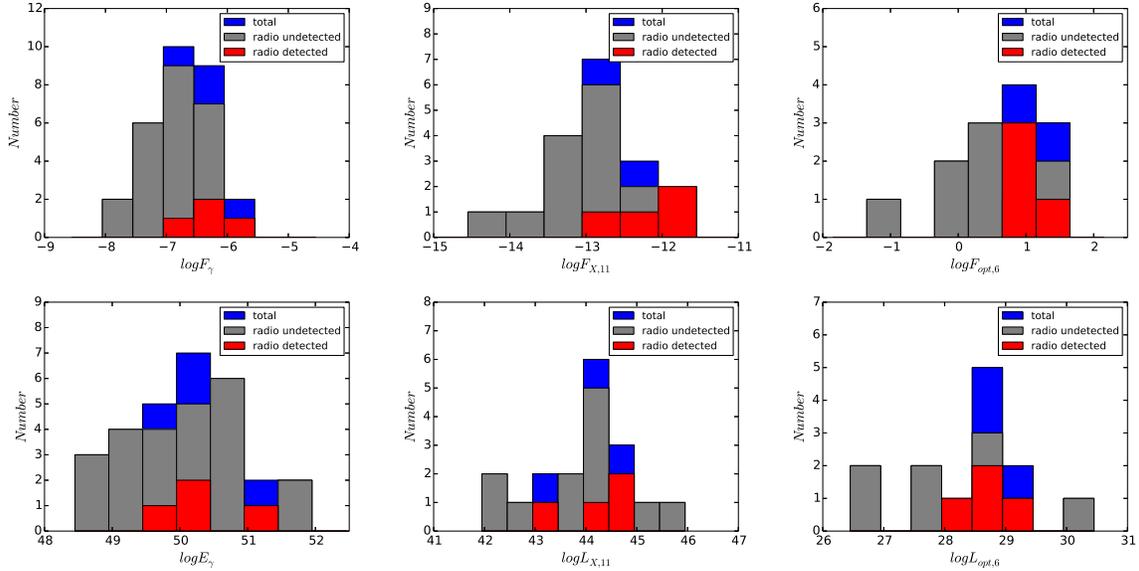}
\end{center}
	\caption{Distributions of $F_{\gamma}$ ,$F_{\rm X,11}$, $F_{\rm opt,6}$ and $E_{\gamma}$, $L_{\rm X,11}$, $L_{\rm opt,6}$. Red and gray color represent samples of GRBs with and without radio afterglow detections and blue represent total sample.}
\label{fig:distributions}
\end{figure*}

Except GRB140903A there are only three SGRBs whose radio afterglows have been detected. The properties of those SGRBs have been summarised by \citet{Berger2014ARA&A}. Although the sample is small, we still can do some primary statistics work in purpose of finding what is the difference between SGRBs with and without radio afterglow detections.

\citet{Berger2014ARA&A} described a sample of 70 SGRBs events with fluence measurements in $15-150$ keV ($F_{\gamma}$), 28(23) of the 70 events have flux of $0.3-10$ keV measurements(up limits) at 11 hours postburst ($F_{\rm X,11}$)  and 28 of 70 events have measured redshifts($ z $).

Together with GRB 140903A which have measured all the $F_{\gamma}$ \citep{Palmer2014}, $F_{\rm X,11}$ and $ z $\citep{Cucchiara2014}, the total sample consists 71 SGRB events. 13 of 71 events have optical afterglow \citep{Fong2015} data which can be fitted by broken power law. Then we can get optical flux density at 6 hours postburst($F_{\rm opt,6}$).

Fig.\ref{fig:relations} shows the relations between fluence and flux (densities) of different bands and distribution of redshifts. We can see that in the three panel at the front, GRBs with radio afterglow detection generally locate in the upper right side and in the last panel they all gather at lower redshift region. This is natural since usually the ``nearby" events  are brighter and easier to be detected (in radio and also other bands).

% which means that SGRBs with detected radio afterglows have smaller redshifts and higher flux or flux densities in other bands.

In order to compare the intrinsic properties of the SGRBs with and without radio afterglow detection, we compare the distributions of $F_{\gamma}$, $F_{\rm X,11}$, $F_{\rm opt,6}$ and the corresponding energy($E_{\gamma}$) and luminosity($L_{\rm X,11}$  and $L_{\rm opt,6}$) in Fig.\ref{fig:distributions}. Those SGRBs with radio afterglow detections have higher values of $F_{\gamma}$, $F_{\rm X,11}$, $F_{\rm opt,6}$ but normal $E_{\gamma}$, $L_{\rm X,11}$  and $L_{\rm opt,6}$. We also note that the radio observations are 
consistent with emission from the forward shock. This means that nearer and brighter SGRBs are more likely to have detectable radio afterglow emission. It is a natural consequence of the limited sensitivity of radio facilities and is consistent with the conclusion made by \citet{Chandra2012} from long GRBs.

\section{Conclusion}

GRB 140903A is characterized by its $\sim 10^{6}$ s radio afterglow emission. Some optical and X-ray afterglow data had also been recorded, rendering GRB 140903A as one of the few short events with afterglow emission in a very wide energy range (i.e., from radio to X-ray).  We show in this work that the afterglow data can be self-consistently reproduced within the forward shock radiation scenario and at $t\leq 7\times 10^{3}$ s an energy injection from the prolonged activity of the central engine is needed. The energy injection form suggests that the central engine is likely millisecond magnetar, similar to that needed for interpreting the X-ray flat segment of SGRB 051221A \citep{Fan2006}. Such a result may suggest a relatively small total-gravitational-mass of the pre-merger binary system or a very hard equation of state of neutron stars that can yield the maximum gravitational mass $>2.3-2.4~M_\odot$, this is because for the ten neutron-star binary systems observed in the Galaxy, their merger remnants will have a typical gravitational mass $\sim 2.3-2.4~M_\odot$, if no significant amount of material has been ejected \citep{Fan2013}. The inferred jet half-opening angle $\sim 0.05~{\rm rad}$ ($2.9^\circ$) is very small, implying that hundreds more similar events took place in the local universe but in directions missing the earth.

%Different from the highly-beamed electromagnetic radiation generated by the ultra-relativistic SGRB outflows, the beam effect of the gravitational wave radiation is weak and hence the expected detection rate of gravitational wave radiation is significantly larger than that of the highly-beamed local sGRBs.

Relative to the ``frequent" detection of SGRB afterglow in X-rays and optical, the radio afterglow emission had just been detected in four events.
To better understand the rare detection in radio bands, we carry out a statistical study to see whether there is a significant difference between the events with and without radio detections. The common features of these radio-detected GRBs are the relatively small redshifts and higher flux or flux density especially in optical band, implying that just the relatively nearby and energetic outflows are easier to give rise to detectable radio emission, consistent with the forward shock afterglow model.

\section*{Acknowledgments}

We thank Dr. Y. Z. Fan for stimulating discussion. This work was supported in part by 973 Programme of China under grant 2014CB845800, National Natural Science Foundation of China under grants 11273063,  11361140349, 11433009, 11303098 and 11103084, the Chinese Academy of Sciences via the Strategic Priority Research Program (Grant No. XDB09000000) and the External Cooperation Program of BIC (No. 114332KYSB20160007).

\clearpage

\end{document}